# Feasibility study on optical vortex generation at Shanghai deep ultraviolet free-electron laser

Haixiao Deng

*Shanghai Institute of Applied Physics, Chinese Academy of Sciences, Shanghai, 201800, P. R. China*

Coherent light with orbital angular momentum (OAM) is of great interest, and more recently, OAM light generation by coupling a relativistic electron beam with a Gaussian mode laser pulse at the high harmonics of a helical undulator has been demonstrated experimentally. In this paper, the possibility of delivering coherent OAM light at the 3$^{rd}$ harmonic of the Gaussian mode seed laser was discussed for the Shanghai deep ultraviolet free-electron laser (SDUV-FEL). The considerations on the experiment setup, the expected performance and the possible measurement method are given.

Email: denghaixiao@sinap.ac.cn

As is well known, light carries both linear and angular momentum. One particular application of the linear momentum of light is optical tweezers [1], in which a highly focused laser beam is utilized to provide an attractive or repulsive force, typically pico-Newtons, depending on the refractive index mismatch to physically hold and move microscopic dielectric particles. Simultaneously, the orbital angular momentum (OAM) of light can also be transferred to particles, causing them to spin, and thus provide the conceptual leap in both understanding and application. In fact, after the breakthrough production of laser beams carrying defined amounts of OAM in the laboratory [2], many branches of optical physics, including nonlinear and quantum optics, information encoding and, of course, optical manipulation has been significantly impacted [3]. Undoubtedly, continuing advances in optical and other technologies will continue to drive forwards OAM of light science and its applications.

In the last two decades, particle accelerator based light sources, e.g., free-electron lasers (FELs) have being developed worldwide to satisfy the dramatically growing demands of high brightness and short-wavelength radiation, especially x-ray pulses which enable the simultaneous probing of both the ultra-small and the ultra-fast worlds. The successful user operations of FEL facilities [4-8] in x-ray regimes have announced the birth of the x-ray laser. Considering its flexible and reliable properties, OAM of light becomes a new topic for FEL community [9-11], besides ultra-short pulse [12-15], full coherence [16-21] and fast polarization switch [22-25]. More recently, using the electron beam as a mode-convertor at the SLAC's next linear collider test accelerator, a Gaussian laser has been experimentally converted to an OAM mode in an 800nm FEL [26], which pave the way for the coherent OAM x-ray pulse with unprecedented brightness.

The Shanghai deep ultraviolet FEL (SDUV-FEL) is a miscellaneous accelerator test facility for x-ray FEL principles and technologies [27], such as echo-enabled harmonic generation lasing [28], two-stage cascaded high-gain harmonic generation [29] and cross undulator [30], etc. It is found that SDUV-FEL is well suited for the lasing experiment of coherent OAM light with some minor modifications. In this paper, the possibility of delivering coherent OAM light at the 3$^{rd}$ harmonic of the 1047nm Gaussian mode seed laser is explored for the SDUV-FEL, and the plan for the experiment is described with the expected performance outlined.

In contrast to a normal seeded FEL scheme, the OAM light generation relies on a harmonic interaction between the electron beam and a Gaussian seed pulse in a circularly polarized undulator to naturally produce a helical energy modulation in the electron distribution, as shown in Fig. 1. After passing through the dispersive chicane, the energy modulation is then converted into the helical micro-bunching, and thus the screw-distribution electrons emit and amplify coherent OAM light in a downstream planar undulator.

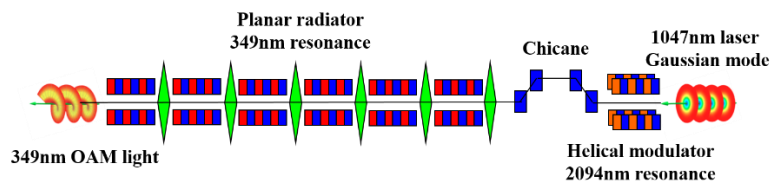

FIG. 1. Schematic setup for OAM light generation at SDUV-FEL.

In current setup of SDUV-FEL, there is no available helical undulator. The helical modulator utilized here is supposed to be a five periods' Apple-II type undulator, as shown in Fig. 2, which is now served as the high power terahertz emitter with the femtosecond electron bunch [31]. The magnetic structure consists of four standard Halbach magnet rows above and below the electron orbit plane, and it can be operated in linear, elliptical or circular polarized modes. The upper-front and lower-back magnet arrays can be moved independently along the longitudinal direction within a range of 60mm. The resonant wavelength of the Apple-II modulator can be calculated by the undulator equation $\lambda_1=\lambda_u(1 + K^2)/2\gamma^2$ for circular polarization, where $\gamma$ is the electron energy normalized to the rest energy $mc^2$, $\lambda_u$ is the undulator period, and $K$ is the undulator parameter defined as $K=eB_0\lambda_u/2\pi mc^2$, with $B_0$ being the peak magnetic field of the undulator. With $B_0$=0.15T and $\lambda_u$=0.1m, Fig. 2 also illustrates the orbit of an electron with beam energy of 135MeV, and the resonant wavelength of the helical modulator is 2094nm in this case.

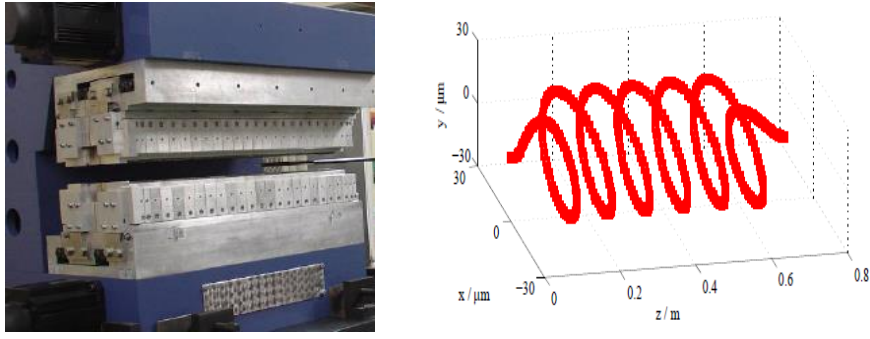

FIG. 2. The Apple-II undulator and the helical orbit of 135MeV electron beam.

In such seeded FEL experimental configurations, the initial uniform electron beam interacts with the 1047nm Gaussian mode laser pulse in the helical modulator. As a consequence of three-dimensional harmonic interaction, the linearly polarized seed laser field profile $E(r)/E_0=\exp(-r^2/w^2)$ excites a helical modulation via the second harmonic resonance. The coupling is proportional to the radial variation of the Gaussian laser field, thus the electrons on-axis are unmodulated, while those at the helical position $ks-h\varphi = n\pi$ and the radial position $r = w/2^{0.5}$ receive the largest energy kick, given by

$$\Delta\gamma_{max} \cong \frac{eK^2N\lambda_u^2 E_0}{10\pi\gamma^2 mc^2 w}, \quad (1)$$

where $N$ is the modulator periods number, $s$ is the electron longitudinal position, $e$ is the charge of an electron, $k$ is the wave number of the seed laser, $h$ and $\varphi$ is the azimuthal mode and angle, respectively. The following dispersive magnetic chicane, characterized by the matrix transport element $R_{56}$ converts the energy modulation into a helical bunching. At the chicane exit, the bunching factor corresponding to the $n^{th}$ harmonic of the seed laser and the $h^{th}$ azimuthal mode, is then can be expressed as

$$b(n,h) = |< e^{-inkz-ih\varphi} >|. \quad (2)$$

In order to ensure that the OAM light dominates in the subsequent radiator, the correlated helical structure must exceed the intrinsic shot noise of the electron beam. Thus, a three-dimensional universal algorithm [32] is used for numerically optimizing the laser-beam interactions in the helical modulator. In the simulation, both the seed laser (1047nm, 2MW) and the electron beam (2μmrad emittance, 100A peak current, 5keV sliced energy spread) are focused to 250μm.

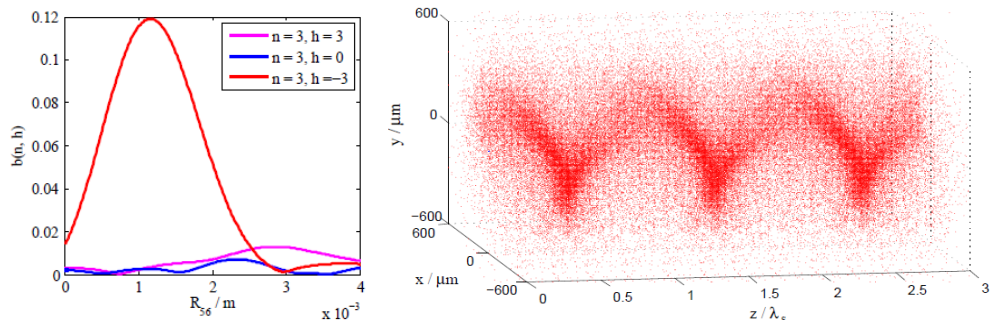

FIG. 3: The bunching factor and the optimized beam distribution after the dispersive chicane.

Fig. 3(a) shows the bunching factor dependences on the chicane dispersion $R_{56}$, when a field strength of $E_0$=56MV/m at the laser waist in the modulator yields an energy modulation amplitude of 25keV. For the 3$^{rd}$ harmonic of the seed laser, $h$=-3 OAM light shows the maximum bunching factor about 12% while other azimuthal modes are well controlled. In the case of the optimal dispersion $R_{56}$=1.2mm, Fig. 3(b) illustrates the beam distribution after density modulation, in which a left-handed helical density modulation is clearly observed. One should note that there also exists strong bunching factor at lower harmonics of the seed laser, e.g., $n$=1, $h$=-1 and $n$=2, $h$=-2, however, which is not under consideration here because that the radiator is chosen to be resonant at the 3$^{rd}$ harmonic.

With the optimized parameter above, the helically micro-bunched beam is imported to GENESIS [33] for the OAM light lasing calculation and the main radiator is tuned to the 3$^{rd}$ harmonic of the seed laser, i.e., 349nm. The main radiator consists of 6 segments of 1.5m planar undulator with undulator period of 25mm and undulator parameter of 1.42. The 12% coherent bunching factor of the $h$=-3 OAM mode is significantly larger than the intrinsic shot noise in modern FELs, and thus it is expected that the $h$=-3 OAM mode will be dominant at the entrance of the radiator, and will be amplified further. The simulated 349nm radiation evolution is then given in Fig. 4, where the radiation peak power exceeds 10MW and saturates after 5 segments of the radiator.

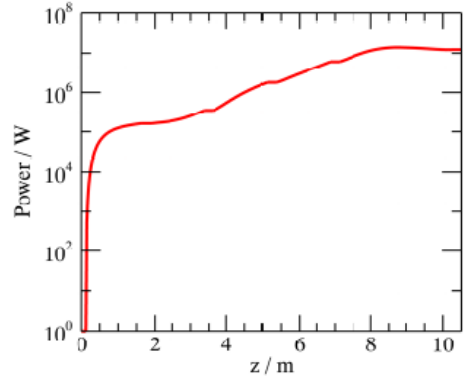

FIG. 4: The 349nm radiation power growth in the main radiator undulator.

Now we consider the azimuthal mode of the 349nm radiation in the radiator. Fig. 5(a) illustrates the distribution of the radiation intensity and phase after the first ten periods of radiator. It is clear from the radiation phase that the large bunching factor of $h$=-3 is responsible for the initial production of the coherent OAM light with $h$=-3. The imperfect roundness in the simulated intensity patterns indicates that other optical modes also exist in the 349nm radiation, which is mainly due to the asymmetries of the electron beam, the impurity OAM mode excitation in the helical modulator [26].

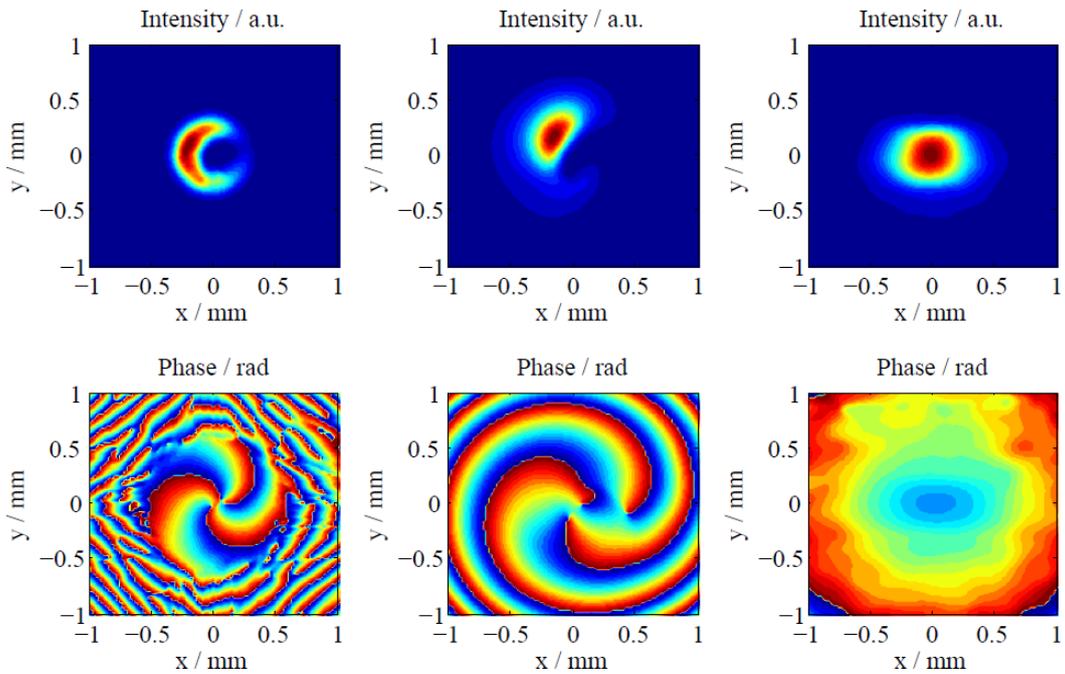

FIG. 5: The azimuthal distribution of 349nm radiation in the radiator undulator. (a) at the position of $z$=0.25m, (b) at the position of $z$=1.5m and (c) at the position of $z$=6.0m.

Fig. 5(b) and 5(c) plots the azimuthal performances of the 349nm radiation after the first segment of the radiator and after saturation, respectively. With the increase of the radiator undulator length, the $h=-3$ OAM mode was surpassed by the $h=0$ Gaussian mode, which is due to the shorter gain length of the Gaussian mode than that of the OAM mode. It arises us an important issue for the OAM mode lasing of a FEL source, besides the initial helical density modulation, i.e., the gain length of the interested OAM mode should be short enough to ensure that it dominates in the radiator undulator with respect to other optical modes.

In Ref. 26, the OAM radiation was confirmed experimentally from the diffraction pattern when passing through a slit and an iterative phase reconstruction procedure from the known transverse profiles of radiation, which may be utilized in the experiment of SDUV-FEL. Meanwhile, considering the cross-correlation method widely used for characterizing FEL properties [34], here we propose and investigate a more straightforward way on the basis of observing cross-correlations of the radiation field in the transverse plane. The cross-correlation intensity are defined using the following expression

$$\xi(x,y) = |E(x,y) + E(x,-y)|^2. \qquad (3)$$

Under such circumstance, one of the arms should be the mirror image of the FEL radiation in cross-correlation. Using the simulated radiation mentioned above, Figure 6 illustrates a set of cross-correlation intensities for three different position of the radiator, z =0.25, 1.5 and 6m respectively. It can be readily seen that there is an obviously different transverse pattern between the Gaussian mode and OAM light. In addition, the six petals in a $2\pi$ angular period confirms the fact of an OAM FEL radiation with an absolute topological charge of $h=3$.

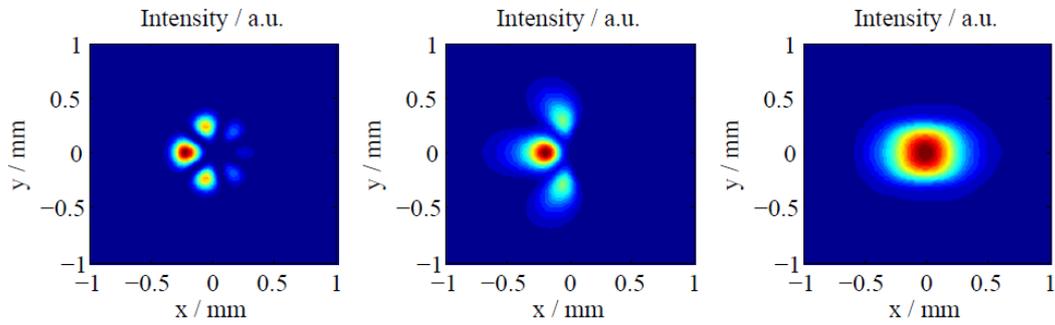

FIG. 6: The cross-correlation of 349nm radiation at the undulator position of 0.25m, 1.5m and 6.0m respectively.

In summary, coherent OAM light generation from a relativistic electron beam has been recently demonstrated in the accelerator lab, by transforming the electron beam distribution to a helix by a Gaussian mode laser. In this paper, together with a 3$^{rd}$ harmonic up-conversion, one possible situation of an OAM lasing experiment at SDUV-FEL is presented. The preliminary considerations on the experiment setup and the expected performance are discussed. Moreover, we propose a cross-correlation method for OAM light characterization. It is shown that the $h=-3$ OAM light at the 3$^{rd}$ harmonic of the 1047nm Gaussian laser can be effectively generated. It is worth stressing that the calculation here is not fully optimized and there is still room for improvement, e.g., the helical modulation and the electron beam size in the radiator could be further matched, and the LINAC beam dynamics could be optimized for the interested OAM light, rather than a Gaussian mode. Meanwhile, we expect that OAM light approach can be extended to the short-wavelength FELs, with the most recent seeding schemes [11, 20-21] and helical undulator driven harmonic lasing scheme of an FEL oscillator [35].

The author would like to thank Jun Yan from Duke University for enthusiastic discussions on the helical modulation calculations, Chao Feng, Tong Zhang, Bo Liu and Dong Wang for FEL physics. This work was partially supported by the Major State Basic Research Development Program of China (2011CB808300) and the National Natural Science Foundation of China (11175240, 11205234 and 11322550).